\documentclass[floatfix,twocolumn]{revtex4}
\usepackage{amsmath}
\usepackage{color}
\usepackage{amssymb}

\usepackage{graphicx}

\begin{document}

\title{{\bf Nuclear spin-lattice relaxation time in UCoGe  }}

\author{V.P.Mineev$^{1,2}$}
\affiliation{$^1$Universite Grenoble Alpes, CEA, IRIG, PHELIQS, F-38000 Grenoble, France\\
$^2$Landau Institute for Theoretical Physics, 142432 Chernogolovka, Russia}

\begin{abstract}
The NMR measurements performed on a single orthorhombic crystal  of superconducting ferromagnet UCoGe ({\it Y.Ihara  et al, Phys. Rev. Lett. {\bf 105} 206403 (2010)})
 demonstrate strongly anisotropic magnetic properties of this material. The presented calculations allow to establish the dependence of longitudinal spin-lattice relaxation rate  from temperature and magnetic field. The  value $ (1/T_1T)$  in field perpendicular to spontaneous magnetisation directed along $c$-axis has maximum in vicinity of Curie temperature
whereas it does not reveal similar behaviour in field parallel to the direction of spontaneous magnetisation. Also there was shown that the longitudinal spin-lattice relaxation rate is strongly field dependent when the field directed in $b$-crystallographic direction but field independent if magnetic field is oriented along $a$-axis.
\end{abstract}

\date{\today}
\maketitle
\section{Introduction}
The uranium ferromagnetic superconductors UGe$_2$, URhGe and UCoGe discovered more than decade ago still continue
attract attention of condensed matter community.
The  superconducting properties of these materials originate from the unusual pairing mechanism induced by magnetic fluctuations (see the recent experimental \cite{Flouquet2019} and theoretical \cite{Mineev2016} reviews and references therein).

Commonly these compounds are considered  as itinerant f-electron metals meaning that the magnetism is determined by the band electrons according to the Stoner mechanism. Indeed, the numerical calculations show that
the contribution  of f-electrons to the bands intersecting the Fermi level is significant. 
However, the x-ray magnetic circular dichroism  measurements \cite{Taupin2015}   and band structure calculations \cite{Samsel2010} point to the local nature of the ferromagnetism in UCoGe. Namely, the comparison of the total uranium moment $ M^{U}_{tot }$ to the total magnetisation $ M_{tot}$ at different magnitude and direction of magnetic field  indicates that the magnetic moment of cobalt ions and   itinerant electrons put together insignificant part   of total magnetic moment, hence the uranium ions dominate the magnetism of UCoGe. The same is true also in the related compounds URhGe \cite{Sanchez2017} and
UGe$_2$ \cite{Kernavanois2001}. 

In all these compounds the magnetic moment per atom at zero temperature $M_0 $ is  smaller than the magnetic moment per atom $M_{CW}$
determined from Curie-Weiss law in paramagnetic state. According to the Wohlfarth criteria  the degree of itineracy increases with increasing the difference $M_{CW}-M_0 $. In  the  present compounds the difference $M_{CW}-M_0 $ characterises the degree of itineracy of f-electrons, but not  an itinerant character of ferromagnetic state.
In the band language the local nature of magnetism  means that the magnetic moment   is furnished by  electrons in spin-up and spin-down states with different orbital momentum projection  filling the cellar below Fermi level \cite{Lander1991,Shick2004,Samsel2010}.  In the real space these states looks like the f-type     Wannier states. 
The same is true in case of d-electrons in transient metals \cite{Eriksson1990,Chen1995}. This type of magnetism  is different from the Heisenberg ferromagnetism of isolated magnetic moments of magnetic ions formed by the electron states split by the crystal field as well from the pure Stoner-Hubbard magnetism of itinerant electrons. A microscopic description of such type magnetic state is not developed.

 The  nuclear magnetic resonance (NMR) 
is one of the main tools for study   magnetic properties of metals in normal and superconducting state. 
The  direction-dependent $^{59}$Co NMR measurements \cite{Ihara2010} of the Knight-shift and nuclear spin-lattice relaxation time provide  knowledge 
of static and dynamic susceptibility components in this orthorhombic metal. The relaxation rate in UCoGe in field perpendicular to the easy magnetisation axis along $c$-crystallographic direction strongly surpasses the relaxation rate in field  parallel to it.
Moreover, there was demonstrated that in vicinity of FM transition the spin-lattice relaxation rate in magnetic field along the easy magnetisation is practically  not deviated from Korringa type behaviour, while  in the other field directions
 there is strong enhancement of relaxation (see Fig.1).
The NMR and $^{59}$Co nuclear quadrupole resonance studies on YCoGe compound free of uranium f-electrons shows that d-electron of Co atoms play no role in ferromagnetism of UCoGe originating from the U-5f electrons
\cite{Ihara2010,Karube2011}.

 Here we present calculations allowing  to establish the peculiar dependence of longitudinal spin-lattice relaxation rate from temperature and magnetic field. The paper is organised as follows.
 As it was explained above the magnetism of UCoGe cannot be described in frame a model of pure itinerant metallic state. Nevertheless we find pertinent to begin with reminder of the spin-lattice relaxation rate derived by T.Moriya and K.Ueda \cite{Moriya1973,Moriya1985} in  the isotropic itinerant ferromagnets. It is
 expressed through  the static  susceptibility along the external field. Thus, it is independent from field direction what is evidently inapplicable to NMR in UCoGe.

Then, to interpret NMR 
observations one need to know the wave-vector - frequency dependent  magnetic susceptibility tensor in the orthorhombic UCoGe compound
that is currently not available. However, the wave vector dependence of static susceptibility components has been established from phenomenological consideration \cite{Mineev2016}. One can obtain the frequency dependence of susceptibility by simple generalisation of corresponding static expression. This allows us to calculate
the NMR longitudinal relaxation rate $1/T_1$ and establish its temperature and field dependence.

 \section{ Nuclear magnetic resonance  relaxation rate}
 
 \subsection{$1/T_1$ in isotropic itinerant ferromagnets}
 
 Before the discussion the anisotropic NMR properties of UCoGe it is pertinent  remind first the theoretical results obtained  by T.Moriya and K.Ueda \cite{Moriya1973,Moriya1985} in the application to the isotropic itinerant ferromagnets.
 In this case
nuclear spin-lattice relaxation rate measured in a field  H along the spontaneous magnetisation chosen as $z$-direction is determined by correlation function of field fluctuations
in the direction perpendicular to the external  field. The latter according to the fluctuation-dissipation theorem is expressed trough the imaginary part of susceptibility $\chi_{+-}^{\prime\prime}({\bf k},\omega)$ of electron gas in the direction perpendicular to the external field
\begin{equation}
\left(\frac{1}{T_{1}T}\right)_z
\propto \lim_{\omega\to 0}\int\frac{d^3k}{(2\pi)^3}
 |A_{hf}|^2\frac{\chi_{+-}^{\prime\prime}({\bf k},\omega)}{\omega}.
\end{equation}
The calculations have been  done in the paper \cite{Moriya1973} yield 
\begin{eqnarray}
\left(\frac{1}{T_{1}T}\right)_z
\propto 
\left\{
 \begin{array} {rc}
\chi_{z}(T)~~~~~~- \mbox{\text paramagnetic ~state}
,\\
~~\\
M^{-2}~~~~~~-\mbox{\text ferromagnetic ~state},
 \end{array} \right.
 \label{chi}
 \end{eqnarray}
where $ \chi_{z}(T)$ is the temperature dependent static susceptibility in the field direction and $M=M(T)$ is the magnetisation in ferromagnetic state. 
Thus,
the summation of perpendicular susceptibility  over reciprocal space gives rise the longitudinal susceptibility !
The origin of this result  is that $\chi_{+-}({\bf k},\omega)$ is determined through the spin-up spin-down band splitting expressed through the static longitudinal susceptibility in paramagnetic state and through the $M^{-2}$ in ferromagnetic state. 
One can find modification of this result in finite external magnetic field in the paper \cite{Kontani1976}, where in particular shown that 
\begin{equation}
\lim_{H\to \infty}\left(\frac{1}{T_{1}T}\right)_z
\to 0
\end{equation}
as it should be in presence of the strong band splitting.

\bigskip

Another important point essential for results Eq.(\ref{chi}) is that the $\chi_{+-}^{\prime\prime}({\bf k},\omega) \propto\omega/kv_F$ both in the noninteracting and interacting Fermi gas such that the integration over reciprocal space in Eq.(1) is in fact two-dimensional.

\subsection{$1/T_1$ in UCoGe}

In the orthorhombic metal the nuclear spin-lattice relaxation rate $1/T_1$ measured in a field  along the $l$ direction is expressed in terms of the imaginary part of the dynamic susceptibility along 
the $m$ and $n$ directions, perpendicular to $l $, $\chi^{\prime\prime}_{m,n}({\bf k},\omega)$ as
\begin{equation}
\left(\frac{1}{T_1T}\right)_l
\propto \sum_{\bf k}
\left[ |A_{hf}^m|^2\frac{\chi^{\prime\prime}_m({\bf k},\omega)}{\omega}+  
 |A_{hf}^n|^2\frac{\chi^{\prime\prime}_n({\bf k},\omega)}{\omega}
\right ].
\end{equation}
At low temperatures the relaxation rate $(1/T_1T)$ for ${\bf 
H}$ parallel to  $c$ crystallographic direction is more than order magnitude smaller  those measured in the other two field directions  parallel to  $a$ and $b$ axes \cite{Ihara2010} (see Fig.1).
So, one can use the expressions for 
\begin{eqnarray}
\left(\frac{1}{T_1T}\right)_b
\propto \sum_{\bf k}
 |A_{hf}^c|^2\frac{\chi^{\prime\prime}_c({\bf k},\omega,H_b)}{\omega}
, 
\label{5}\\
\left(\frac{1}{T_1T}\right)_a
\propto \sum_{\bf k}
 |A_{hf}^c|^2\frac{\chi^{\prime\prime}_c({\bf k},\omega,H_a)}{\omega}
,
\label{6}
\end{eqnarray}
neglecting the terms, originating from the imaginary part of susceptibilities $\chi_a^{\prime\prime}$ and $\chi_b^{\prime\prime}$.

The static susceptibilities in uranium ferromagnets are derived in the paper \cite{Mineev2016}.  Along the $c$-axis  it  is 
\begin{equation}
\chi_c({\bf k})=\frac {1}{(\chi_c(H_l))^{-1}+2\gamma^c_{ij} k_ik_j},
\end{equation}
where $\chi_c(H_l)=\frac{\partial M_c}{\partial H_c}$ is the homogeneous static susceptibility along $c$-axis at fixed value  of magnetic field  $H_l$ along $l=a,b$
direction.
The simplest generalisation for dynamic case obeying the Kramers-Kronig relation $$\chi_c({\bf k})=\frac{1}{\pi}\int\frac{\chi^{\prime\prime}_c({\bf k},\omega)}{\omega}d\omega$$
is the following 
\begin{equation}
\chi_c({\bf k},\omega)=\frac {1}{-\frac{i\omega}{A}+(\chi_c(H_l))^{-1}+2\gamma^c_{ij} k_ik_j},
\end{equation}
where $\chi^{\prime\prime}_c({\bf k},\omega)=\mbox{\text Im}~\chi_c({\bf k},\omega)$, and  $A$ is a constant.
Thus, to estimate the sum over momenta we can  work with 
\begin{equation}
\frac{\chi^{\prime\prime}_c({\bf k},\omega,H_l)}{\omega}=\frac{A}{\omega^2+A^2\left [ (\chi_c(H_l))^{-1}+2\gamma_{ij} k_ik_j\right ]^2}.
\end{equation}

The field dependence of $\chi_c(H_l)$ can be determined as follows \cite{Mineev2016}.
The Landau expansion of the free energy near the Curie temperature in presence of magnetic field is
\begin{eqnarray}
F=\alpha_cM_c^2+\beta_cM_c^2+\alpha_aM_a^2+\alpha_bM_b^2\nonumber\\
+\beta_{ac}M_a^2M_c^2+\beta_{bc}M_b^2M_c^2 -{\bf H}{\bf M},
\label{F11}
\end{eqnarray}
where 
\begin{equation}
\alpha_c=\alpha_{c0}(T-T{c0}),~~~\alpha_a>0,~~~~~\alpha_b>0.
\end{equation}
In constant magnetic field perpendicular to the spontaneous magnetisation ${\bf H}=H_b\hat b$
the equilibrium magnetisation projection along the $b$ direction
\begin{equation}
M_b\approx\frac{H_b}{2(\alpha_b+\beta_{bc}M_c^2)}
\label{My}
\end{equation} 
is obtained by minimisation of free energy (\ref{F11}) in respect of $M_b$.
Substituting this expression back to (\ref{F11})
we obtain
\begin{eqnarray}
F=\alpha_{c}M_{c}^{2}
+\beta_{c}M_{c}^{4}-\frac{1}{4}\frac{H_b^2}{\alpha_b+\beta_{bc}M_c^2},
\label{F1}
\end{eqnarray}
that gives after expansion of the denominator in the last term, 
\begin{equation}
F=-\frac{H_b^2}{4\alpha_b}+\tilde\alpha_{c}M_{c}^{2}
+\tilde\beta_{c}M_{c}^{4}+\dots,
\label{F2}
\end{equation}
where
\begin{eqnarray}
&\tilde\alpha_{c}=\alpha_{c0}(T-T_{c0})+\frac{\beta_{bc}H_b^2}{4\alpha_b^2},\\
&\tilde\beta_{c}=\beta_c-\frac{\beta_{bc}}{\alpha_b}\frac{\beta_{bc}H_b^2}{4\alpha_b^2}\label{beta}.
\end{eqnarray}
Thus, in a magnetic field ${\bf H}=H_b\hat b$ perpendicular to the direction of spontaneous magnetisation  the Curie temperature decreases as
\begin{equation}
T_c(H_b)=T_{c0}-\frac{\beta_{bc}H_b^2}{4\alpha_b^2\alpha_{c0}}.
\label{Cur b}
\end{equation}
The corresponding formula for field parallel to $a$-axis is
\begin{equation}
T_c=T_c(H_a)=T_{c0}-\frac{\beta_{ac}H_a^2}{4\alpha_a^2\alpha_{c0}}.
\label{Cur a}
\end{equation}
The coefficient $\alpha_a\gg\alpha_b$ that is the $a$-direction is much harder magnetically than the $b$-direction. Hence, the Curie temperature
is practically  independent from magnetic field in $a$-direction $T_c(H_a)\approx T_{c0}$. 
The susceptibility
 along $c$-axis 
\begin{eqnarray}
\chi_c(H_b)=\left \{\begin{array}{l}\frac{1}{4\alpha_{c0}\left ( T_{c0}-\frac{\beta_{bc}H_b^2}{4\alpha_b^2\alpha_{c0}}-T \right )}, ~~~~~~T<T_c(H_b),\\
\frac{1}{2\alpha_{c0}\left (T- T_{c0}+\frac{\beta_{bc}H_b^2}{4\alpha_b^2\alpha_{c0}}\right )},~~~~~T>T_c(H_b) \end{array} \right.
\end{eqnarray}
increases with magnetic field along $b$-axis but keeps the constant value in magnetic field parallel to $a$-axis
\begin{eqnarray}
\chi_c(H_a)=\left \{\begin{array}{l}\frac{1}{4\alpha_{c0}\left ( T_{c0}-T \right )}, ~~~~~~T<T_{c0},\\
\frac{1}{2\alpha_{c0}\left (T- T_{c0}\right )},~~~~~T>T_{co} \end{array} \right..
\end{eqnarray}

Now, let us make the integration in the Eqs. (\ref{5}), (\ref{6}).
For the simplicity one can calculate  the converging  integral in the spherical approximation 
\begin{eqnarray}
\left (\frac{1}{T_1T}\right)_l
\propto 
 \int\frac{4\pi k^2dk}{(2\pi)^3}\frac{A}{\omega^2+A^2\left [ (\chi_c(H_l))^{-1}+2\gamma k^2\right ]^2}\nonumber\\=\frac{\sqrt{2}}{32\pi A\gamma^{3/2}}
 \frac{\sqrt{\chi_c(H_l)}}{\left (1+\frac{\omega^2\chi^2_c(H_l)}{A^2}\right )^{1/4}}\frac{1}{\cos \left (\frac{1}{2} \arctan\frac{\omega \chi_c(H_l)}{A}\right )}.
\end{eqnarray}
We see, that 
\begin{eqnarray}
\frac{1}{T_1(H_b)T}\propto\left \{\begin{array}{l}\sqrt{\chi_c(H_b)},~~~~~\chi_c(H_b)\ll \frac{A}{\omega}\\
~~\\
\sqrt {\frac{A}{\omega}},~~~~~~~~~~~~\chi_c(H_b)\gg\frac{A}{\omega}\end{array} \right.
\end{eqnarray}
 as function of temperature reaches maximum in vicinity of Curie temperature. The same is true for relaxation rate  $\frac{1}{T_1(H_a)T}$ in field parallel to $a$-axis.
 
The magnetic field along {b}-axis shifts the maximum of relaxation rate  to the lower temperature, the  field along $a$-axis does not. This effect is clearly demonstrated by the measurements 
in high enough fields reported in the paper \cite{Hattory2014}. The deviation of the maximum of $\frac{1}{T_1(H_b)T}$ from the maximum of $\frac{1}{T_1(H_a)T}$  in a weak  field $H=2$ T  on Fig.1 is probably originates from a slight misalignment of field orientation as  pointed in  the paper \cite{Hattory2012}).

The similar calculation of spin-lattice relaxation  in field  parallel to $c$-axis $\frac{1}{T_1(H_c)T}$ is expressed through the magnetic susceptibilities along $a$ and $b$ crystallographic directions.  Both of them are much smaller than susceptibility in $c$-direction and both of them are not changed in temperature interval near the Curie temperature. Hence, $\frac{1}{T_1(H_c)T}$ in vicinity of critical temperature is practically temperature independent.

\section{Conclusion}

We have demonstrated  that in contrast with the isotropic weak ferromagnets the longitudinal spin-lattice relaxation rate in UCoGe is expressed through the 
static susceptibility in the perpendicular to magnetic field direction. The  value $ (1/T_1T)$  in field perpendicular to spontaneous magnetization has maximum in vicinity of Curie temperature
whereas it does not reveal similar behaviour in field parallel to the direction of spontaneous magnetisation. These results are in qualitative correspondence with
experimental data presented in Fig.1 \cite{Ihara2010}.

Also there was shown that the longitudinal spin-lattice relaxation rate is strongly field dependent when the field directed in $b$-crystallographic direction but field independent if magnetic field is oriented along $a$-axis what also  accords with experimental observations \cite{Hattory2014}.

The presented calculations have been done using isotropic wave vector dispersion of susceptibilities. The real dispersion laws can be quite different and the simple relationship between the relaxation rate and the susceptibility will be lost. However, the  field-temperature dependence of $1/T_1$ will qualitatively remain.

Another observation which possibly can be useful for the future theoretical efforts is that the calculations were performed taking the kinetic coefficient $A=$const. 
  However, taking 
$
A\sim kv, 
\label{A}
$
where $v$ is a constant with dimensionality of velocity,
the integration over reciprocal space becomes two-dimensional and we come to
$
\frac{1}{T_{1b}T}
\propto \chi_c(T,H_b).
$
A theoretical description of  wave vector - frequency dependence of magnetic susceptibility of actinide intermetallics is still not at hand.

\acknowledgments

I am indebted to Kenji Ishida for the helpful discussions of  his  experimental results.

\begin{figure}
\includegraphics
[height=.8\textheight]
{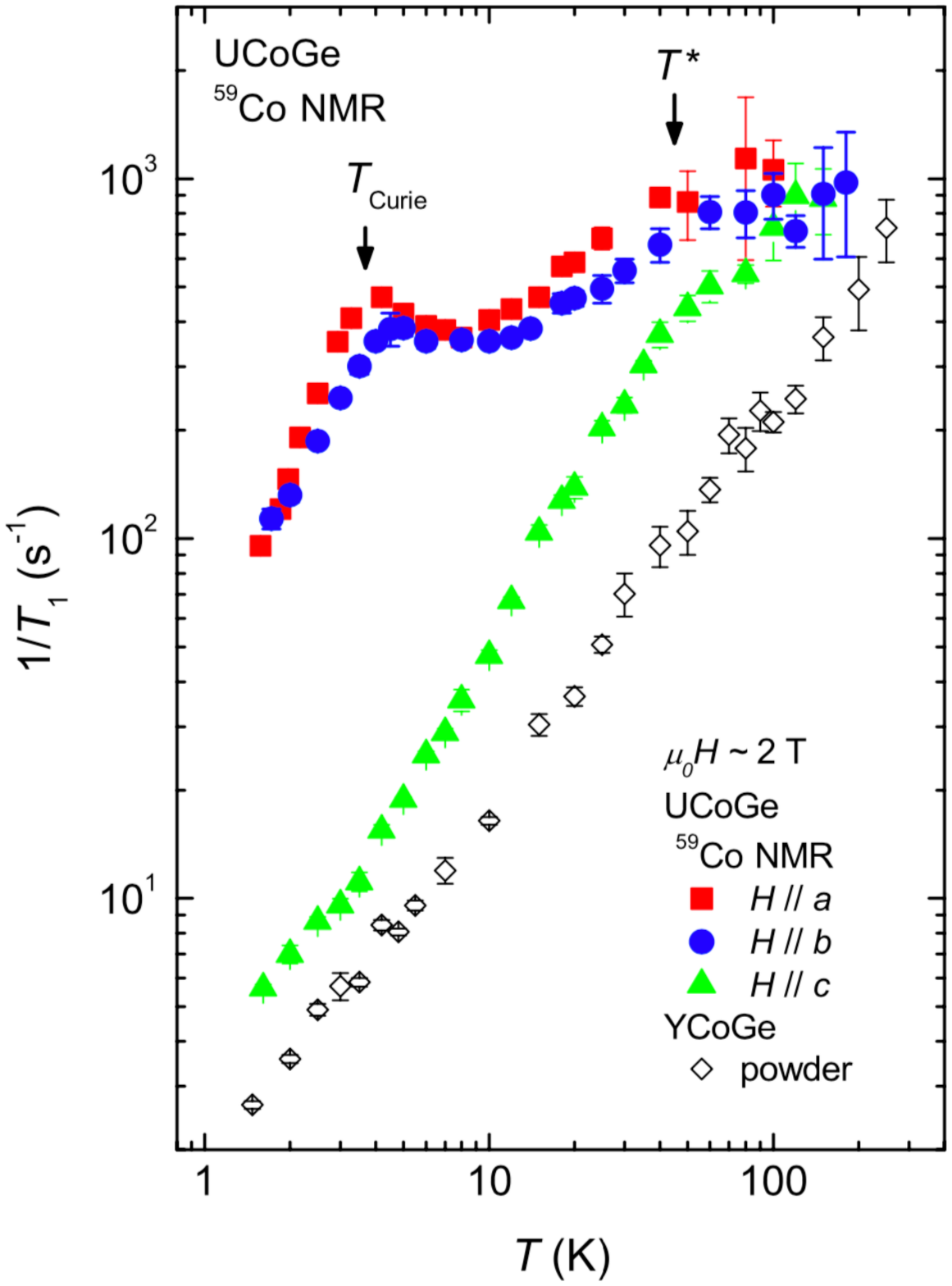}
 \caption{(Color online) 
Nuclear spin-lattice relaxation rate $(1/T_1)$ measured in three different field directions \cite{Ihara2010}. The results of $^{59}$Co NQR on YCoGe, a reference compound without f-electrons, are also displayed.  The Figure is reproduced from the paper Ref.11 with kind permission of prof. K.Ishida.}
\end{figure}


\begin{thebibliography}{220}

\bibitem{Flouquet2019} D.Aoki, K.Ishida and J.Flouquet, J. Phys. Soc. Jpn. {\bf 88}, 022001 (2019).

\bibitem{Mineev2016}V.P.Mineev, Usp. Fiz. Nauk {\bf 187}, 129 (2017) [Phys.-Usp. {\bf 60}, 121 (2017).





\bibitem{Taupin2015} M. Taupin, J.-P. Sanchez,J.-P. Brison,1,D. Aoki, G. Lapertot,F. Wilhelm, and A. Rogalev, Phys.Rev. B {\bf 92}, 035124 (2015).

\bibitem{Samsel2010} M.Samsel-Czekala, S.Elgazzar, P.M.Oppeneer, E.Talik, W.Waleszyk and R.Troc, J. Phys.: Condens.Matter {\bf 22}, 015503 (2010).

\bibitem{Sanchez2017} F.Wilhelm, J.P.Sanchez, J.-P.Brison, D.Aoki, A.B.Shick, and A.Rogalev, Phys.Rev. {\bf 95}, 235147 (2017).

\bibitem{Kernavanois2001}N.Kernavanois, B.Grenier, A.Huxley, E.Ressouche, J.P.Sanchez, and J.Flouquet, Phys. Rev. B {\bf 64}, 174509 (2001).

\bibitem{Lander1991} G.H.Lander, M.S.S.Brooks, B.Johansson, Phys.Rev.B {\bf 43},  13672 (1991).

\bibitem{Shick2004} A.B.Shick, V. Janis, V.Drchal, and W.F.Pickett, Phys.Rev. B {\bf 70}, 134506 (2004).

\bibitem{Eriksson1990}O.Eriksson, B.Johansson, R.C.Albers, A.M.Boring and M.S.S.Brooks, Phys.Rev. B {\bf 42}, 2707 (1990).

\bibitem{Chen1995} C.T.Chen, Y.U.Idzerda, H.-J.Lin, N.V.Smith, G.Meigs, E.Chaban, G.H.Ho, E.Pellegrin, and F.Sette, Phys. Rev. Lett. {\bf 75}, 152 (1995).

\bibitem{Ihara2010}Y.Ihara , T.Hattory,  K.Ishida,  Y.Nakai ,  E.Osaki,
 K.Deguchi, N.K.Sato, and I.Satoh, Phys. Rev. Lett. {\bf 105} 206403 (2010).
 
 \bibitem{Karube2011} K.Karube, T.Hattori, Y.Ihara, Y.Nakai, K.Ishida, N.Tamura, K.Deguchi, N.K.Sato, and H.Harima, J. Phys. Soc. Jpn. {\bf 80},
064711 (2011).


\bibitem{Moriya1973} T.Moriya, K.Ueda, Solid State Commun. {\bf 15}, 169 (1974).

\bibitem{Moriya1985} T.Moriya "Spin fluctuations in itinerant electron magnetism", Springer-Verlag, Berlin, 1985.

\bibitem{Kontani1976} M.Kontani, T.Hioki and Y.Masuda,  Solid State Commun. {\bf 18}, 1251 (1976).

\bibitem{Hattory2014}T.Hattori, K.Karube, K.Ishida, K.Deguchi, N.K. Sato, and T.Yamamura, J. Phys. Soc. Jpn. {\bf 83}, 073708 (2014).



\bibitem{Hattory2012}T. Hattori, Y. Ihara, Y. Nakai, K.Ishida,Y. Tada,S. Fujimoto, N. Kawakami, E. Osaki, K.Deguchi, N. K. Sato, and I. Satoh,
Phys. Rev. Lett. {\bf 108}, 066403 (2012).












\end{thebibliography}
\end{document}